\documentclass[12pt]{article}
\usepackage{amscd,amsmath,amssymb,amsfonts, eucal,
         eurosym, latexsym, mathrsfs,url, verbatim,
        graphicx,subfigure}
        \usepackage{afterpage,lscape,pdflscape,rotating}
\begin{document}
\begin{center}
{\large Nonparametric Clustering of Mixed Data\\ Using Modified
Chi-square Tests}
\end{center}
\vskip 0.1in
\begin{center}
{Yawen Xu,  Xin Gao and Xiaogang Wang\\
Department of Mathematics and Statistics\\
York University, Toronto, Canada}
\end{center}
\vskip 0.1in

\vskip 0.2in
\begin{abstract}
We propose a non-parametric   method to  cluster mixed data
containing both continuous and discrete random variables. The
product space of continuous and categorical sample spaces is
approximated locally by analyzing neighborhoods with cluster
patterns. Detection of cluster patterns on the product space is
determined by using a  modified Chi-square test. The proposed method
does not impose a global distance function which could be difficult
to specify in practice. Results from simulation studies have shown
that our proposed methods out-performed the benchmark method,
AutoClass, for various settings.
\end{abstract}


\section{Introduction}

Mixed data are abundant in scientific research especially in medical
or biological studies. An effective clustering method for mixed data
will  partition a large and complex data set into manageable and
homogeneous subgroups. It thus has wide range applications in almost
any scientific studies including  financial data, personalized medicine
and scientific studies on climate changes.

Most of the  clustering methods in the literature  have been mainly
focused on numerical data. K-mean algorithm has been widely used in
the industry for a long time. Detailed description and discussions
can be found in Kaufman and Rousseeuw (2005). To capture the
intrinsic geometric properties, a suitable distance function such as
Manhattan distance or Mahoblis distances can be used when the
underlying sample space are believed   to be non-Euclidean. K-mode
algorithm by  Huang (1997) extends   this geometrical approach   to
categorical data.   However, this has   proven to be not very
successful for categorical data as demonstrated in Zhang {\it et.
al} (2005). The geometrical or topological natures of continuous and
categorical sample spaces are intrinsically different since the
first one can be endowed with a differential manifold while the
second one is defined entirely on a lattice   with discontinues
functions.    Even when suitable distance functions are valid for
continuous and discrete portion, a challenging question is   how to
combine the metrics from a continuous and a discrete sample space. A
naive approach is to consider a convex combination of the two
metrics which implies that the product space of continuous and
discrete data can be metrizable in this fashion. The major
difficulty is on how to choose the weights without introducing
significant local or global distortions.

Alternatively, a parametric model based on Gaussian mixture could be
used for continuous data, see Banfield and Raftery (1993). One of
the most prominent methods is by Bradley {\it et. al} (1998) which
can be scaled   to large disk-resident data sets. The number of
clusters and outliers can be handled simultaneously. Fraley and
Raftery (1998) propose to choose the number of clusters
automatically for model-based clustering method.  For clustering
mixed data, the AutoClass method proposed by Cheeseman and Stutz
(1995) is well known and could be considered as the benchmark method
for model-based clustering method in this class.  AutoClass takes a
database containing both real and discrete valued attributes, and
automatically finds the number of clusters and groups automatically.
This method has widely used in NASA and it helped to find infra-red
stars in the IRAS Low Resolution Spectral catalogue and discovery of
classes of proteins.

Instead a parametric model, we propose a non-parametric clustering
method which does not assume a global distance function or any
knowledge of the functional form of the joint probability density
function. The key idea is inspired by the fact that any complicated
manifold is supposed to be ``locally" by a manifold with simpler
structure. For example, it is well known that a differentiable
manifold is homeomphic to $R^m$. For categorical data, we suppose
that a neighborhood on a lattice can be sufficiently characterized
by the Hamming distance. The Hamming distance is widely used in
information and coding theory, see Roman (1992) and Laboulias {\it
et. al} (2002). It only measures how many attributes are different
without any attempt to impose any order on the magnitude of the
observed difference. When the true manifold can be approximated
locally by the product space of two manifolds that adopt either
Euclidean or Hamming distance, a statistical test is designed based
a weighted local Chi-squared test. This idea of local test for
clustering was first proposed by Zhang {\it et. al} (2005) for
categorical data.

This article is organized as follows. The method is proposed in
Section 2. The clustering algorithm is presented in Section 3.
Simulation results are provided in Section 4. Discussions are
provided in Section 5.

\section{Method}
In this section, we introduce mixed sample space, on which we adopt
the Hamming distance and Euclidean distance function to measure the
relative positions of two data points.  We define a HD vector, ED
vector and optimal separation point which are essential component
for the proposed weighted local chi-square test for clustering.

\subsection {Joint Sample Space of Mixed Data}
Now consider a general setup for mixed data where $p$ nominal
categorical attributes and q continuous attributes are of interest.
The $j$th categorical attribute is categorized by $m_{j}$ levels
defined by set $A_{j} = (a_{j1}, \cdots,a_{jm_{j}}), j = 1, \cdots,
p$. The categorical portion of data, $\textbf{X} = (\textbf{X}_{1},
\cdots, \textbf{X}_{n})$ is collected from n subjects, with
$\textbf{X}_{i} = (X^{[1]}_{i}, \cdots, X^{[p]}_{i})^{t}$ being the
vector of the observed states of $p$ attributes for subject $i$. The
categorical sample space, $\Omega _{p}$ is defined as a collection
of all possible p-dimensional vectors of states, namely $\Omega _{p}
= \{(\omega_{1}, \cdots, \omega_{p})^{t} | \omega_{1} \in A_{1},
\cdots, \omega_{p} \in A_{p} \}$. The continuous data, $\textbf{Z} =
(\textbf{Z}_{1}, \cdots, \textbf{Z}_{n})$ is collected from same $n$
subjects, with $\textbf{Z}_{i} = (Z^{[1]}_{i}, \cdots,
Z^{[q]}_{i})^{t}$, being the vector of the observed values of $q$
attributes for subject $i$, where $Z^{j}_{i} \in R$ for $i =
1,\cdots, n$ and $j = 1, \cdots, q$.  The continuous sample space is
defined as $\Omega _{q} = R^{q}$. The mixed data consists of
$\textbf{(X, Z)}$ with overall space $\Omega = \Omega_{p} \otimes
\Omega_{q}$.

\subsection { Distance Vectors}
We use Hamming distance (HD) to measure the relative positions of two
categorical data points and Euclidean distance (ED) to measure the
that of two continuous data points.

To be more specific, for any two positions in the categorical sample
space $\Omega_{p}$, $\textbf{X}_{h} = (\omega^{1}_{h}, \cdots,
\omega^{p}_{h})^{t}$ and $\textbf{X}_{i} = (\omega^{1}_{i}, \cdots,
\omega^{p}_{i})^{t}$, the
 Hamming distance (HD) between $\textbf{X}_{h}$ and $\textbf{X}_{i}$ on the $j$th
attribute is
\begin{equation*}
            d_{j}(\textbf{X}_{h},\textbf{X}_{i}) = \left\{
                      \begin{array}{llll}
                         & \hbox{0} & if &\omega^{[h]}_{h} = \omega^{[h]}_{i}, \\
                         & \hbox{1} & if &\omega^{[h]}_{h} \neq  \omega^{[h]}_{i};
                      \end{array}                    \right.
\end{equation*}
and the distance between the two positions is the sum of the
componentwise distances,
\begin{equation*}
     HD(\textbf{X}_{h},\textbf{X}_{i}) = \sum_{j=1}^{p}
     d_{j}(\textbf{X}_{h},\textbf{X}_{i}).
\end{equation*}
For continuous data, the Euclidean distance (ED) between the two
positions is defined as
\begin{equation}
     ED(\textbf{Z}_{h},\textbf{Z}_{i}) = \sqrt{(Z^{[1]}_{h} - Z^{[1]}_{i})^2 + (Z^{[2]}_{h} - Z^{[2]}_{i})^2 + \cdots + (Z^{[q]}_{h} -
     Z^{[q]}_{i})^2}.
\end{equation}

We now introduce HD Vector and ED Vector.  Let
$(\textbf{S},\textbf{T})$ be a reference position in the sample
space with $S = (s_{1}, \cdots, s_{p})\in R^p$ and $T = (t_{1},
\cdots, t_{q})\in R^q$. We measure the distance of all data points
to the selected reference point.  For the categorical portion,
$HD(X_{i}, S)$ can take values ranging from $0$ to $p$. We define
the HD vector to record the frequencies of each distance value.
Namely, the HD vector $U(\textbf{S})$ is a (p+1)-element vector and
is defined as $(U_{0}(\textbf{S}),
U_{1}(\textbf{S}),...U_{p}(\textbf{S}))^t$ where the $j^{th}$
component is given by
\begin{equation*}
   U_{j}(S) =  \sum_{i=1}^{n} \emph{\textbf{I}}\; [HD(\textbf{X}_{i},\textbf{S}) = j], j = 0, 1, \cdots
   p,
\end{equation*}
where $\textbf{I}(A)$ is the indicator function that takes value $1$
when event $A$ happens and value $0$ when  event $A$ does not
happen. For continuous portion of the data, in order to construct a
frequency vector of $ED(Z_{i}, \textbf{T})$ , we need to choose the
bin size. The choice of bin size should be user defined.  In practice,
we find that choosing bin size $l$ equal to $10$ gives satisfactory empirical
result. Let $\textbf{B} = (B_{1}, \cdots, B_{l})$ denote a set of
equal-sized intervals $[b_j^{\;l}, b_j^{\;u}], j=1,2,\cdots, l$.  An
ED vector is defined as $\textbf{V(T)} = [V_{1}(\textbf{T}),
V_{2}(\textbf{T}),...V_{l}(\textbf{T})]^t$, where the $j$th
component is the frequency given by
\begin{equation*}
   V_{j}(\textbf{T}) = \sum_{i=1}^{n} \emph{\textbf{I}}\; [ED(\textbf{Z}_{i},T) \in B_{j}], \;\;\;\ j = 1,\cdots,
   l.
\end{equation*}

In order to use the HD vector ED vector to detect possible clusters,
we define a reference or null HD vector and ED vector when there is
no clustering pattern in the mixed sample space $\Omega_{p} \otimes
\Omega_{q}$. If there is indeed no pattern, then it is equally
likely for a randomly chosen data point to take any  possible
position in the joint sample space.  The resulting HD vector is
called {\it uniform} HD vector (UHD) and ED vector is called
{\it uniform} ED vector (UED) which record the the expected
frequencies under the null hypothesis that there are no clustering
patterns in data.  Let \textbf{X} be a categorical portion of data
and Z be a continuous portion of the data from a sample of size $n$,
with each observation having an equal probability of locating at any
position on space $\Omega_{p} \otimes \Omega_{q}$.  The expected
value of HD vector and ED vector associated with the null
hypothesis, denoted by $\boldsymbol\varepsilon = ( \varepsilon_{0},
\cdots \varepsilon_{P})^t$ and $\boldsymbol\nu = ( \nu_{1}, \cdots
\nu_{l})^t$ are denoted  as the UHD vector and UED vector,
respectively.

Zhang {\it et. al} (2005) proved that the UHD takes  the form of $\boldsymbol\varepsilon =
\frac{n}{M}  \textbf{U}^{*}$, where $M = \prod_{j=1}^{p} m_j, j =
1,2,\cdots,p$ ; and $\textbf{U}^{*} = (U_{0}^*, U_{1}^*, \cdots,
U_{p}^*)$ with
\begin{equation*}
%\left\{
  \begin{array}{l}
      \hbox{$U_{0}^* = 1$;} \\
      \hbox{$U_{1}^* = (m_{1} -1) + (m_{2} - 1) + \cdots + (m_{p} - 1)$;} \\
      \hbox{$U_{2}^* =  \sum_{i<j}^{p} (m_{i}-1)(m_{j}-1)$;} \\
      \hbox{\vdots} \\
      \hbox{$U_{p}^* = (m_{1} -1)(m_{2} - 1)\cdots (m_{p} - 1)$.}
  \end{array}
%\right.
\end{equation*}
\parindent 20pt
For continuous data, the exact distribution of the UED vector is not tractable. We then obtain
 this vector by computer simulations. We
simulate random data points with $q$ continuous independent
attributes. The UED vector is the sampling  frequencies of
ED vector from the simulations corresponds to the null hypothesis that there are no more than one cluster..
Figure 1 provides the plot of UED vector obtained from simulated
null hypothesis with no clusters and ED vector obtained from a simulated data set with
clustering structures.
\begin{figure}[!ht]
\caption{Plots for ED Vector and UED Vector.}
    \begin{subfigure}
        \centering
        \includegraphics[scale=0.3]{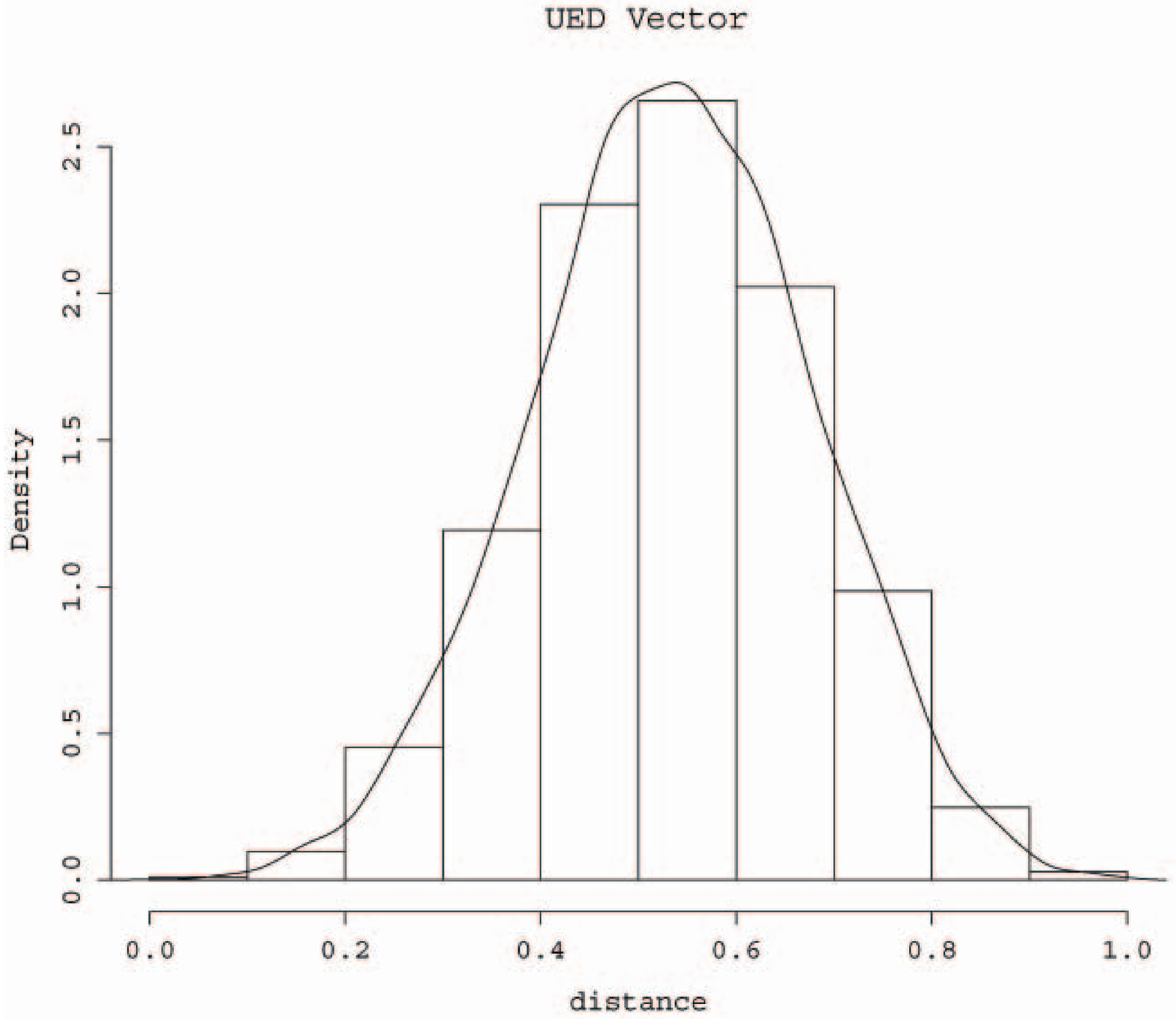}
        %\caption{ED Vector}
    \end{subfigure}
    \begin{subfigure}
        \centering
        \includegraphics[scale=0.3]{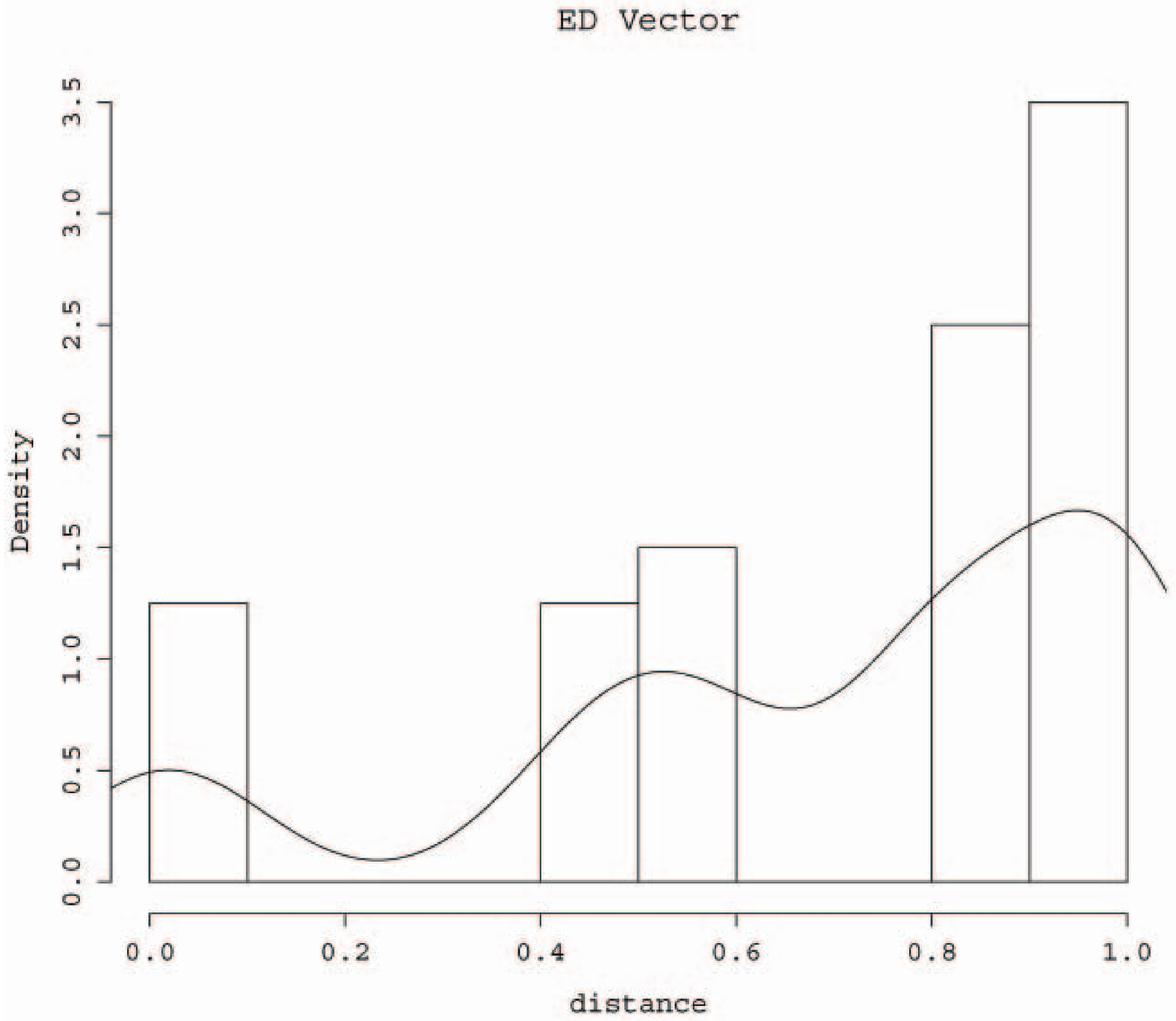}
        %\caption{UED Vector}
    \end{subfigure}
\end{figure}

\subsection {Optimal Separation Point}

If the initial starting point is chosen to be the center of one
particular cluster, then the frequency of ED  should demonstrate a
general decreasing pattern as the ED function records the
frequencies of data points from the center of cluster and outwards.
Small local bumps at the beginning part of  the ED curve are
expected
  if the initial starting point deviate slightly from the
cluster center. Any substantial  reversal of decreasing trend will
produce a valley area on the ED curve as can be seen from the Figure
2.  This might indicate distances that corresponds to boundary
points of the current cluster. The recorded frequencies might
increase when the function records distances from another cluster.
Therefore, the valley area is an ideal place to perform an operation
to separate data points from the current cluster with the rest.

Assume that the categorical data \textbf{X} and continuous data
\textbf{Z} are not uniformly distributed in the sample space $\Omega
_{p} \otimes \Omega_{q}$. Let $U(\textbf{S}) = (U_{0}(\textbf{S}),
U_{1}(\textbf{S}), \cdots, U_{p}(\textbf{S}))^t, \textbf{S} \in
\Omega_{p}$ be the collection of all $(p+1)$-element HD vectors in
the space  $\Omega _{p}$ and $V(\textbf{T}) = (V_{1}(\textbf{T}),
V_{2}(\textbf{T}), \cdots, V_{l}(\textbf{T}))^t, \textbf{T} \in
\Omega_{q}$ be the collection of all $l$-element ED vectors in the
space $\Omega _{q}$, and let $ \boldsymbol\varepsilon =
(\varepsilon_{0}, \varepsilon_{1}, \cdots, \varepsilon_{p})^t$ be
the UHD vector and $\boldsymbol\nu = (\nu_{1}, \nu_{2}, \cdots,
\nu_{l})^t$ be the UED vector defined in above subsection.  For a
given distance value $ j_c, j = 0, 1, \cdots, p$ for categorical
distance values and $j_d, j_d= 1, 2, \cdots, l$ for continuous
distance values, there always exists at least one position
$\textbf{(S,T) } \in \Omega_{p} \otimes \Omega_{q}$, such that the
frequency at this distance value is lager than the corresponding
component, $\varepsilon_{j}$ of the UHD vector
$\boldsymbol\varepsilon$ and $\nu_{j} $ of the UED vector
$\boldsymbol\nu$. In order to compare the HD vector with the UHD
vector, and ED vector to UED vector, we introduce a selection
criterion for an optimal separation or cut-off point $r^*$. The
categorical cut-off was defined and proved by Zhang {\it et. al} (2005). We
extend their approach to the continuous portion of data.  If the clusters
structure is present, the early segment of an HD vector and ED
vector with respect to a data center should contain substantially
larger frequencies than the corresponding frequencies of the UHD
vector and UED vector respectively.
When the observed distances vectors are intersecting and going below the UHD or UED vectors,
valley areas are created and they provide good hints about the locations of optimal
separation points.  This leas to an optimal $r_c^*$ for categorical
portion of data be:
\begin{equation*}
   r_{c}^*(\textbf{S}) = \min_{j_c>0}\{j| \frac{U_{j_c}(\textbf{S})}{\varepsilon_{j_c}} < 1 \}
   -1, \textbf{S} \in \Omega_{p}.
\end{equation*}
Similarly, optimal $r_d^*$ for continuous portion of data be:
\begin{equation*}
   r_{d}^*(\textbf{T}) = \min_{j_d>1}\{j| \frac{V_{j_d}(\textbf{T})}{\nu_{j_d}} < 1
   \}, \textbf{T} \in \Omega_{q}.
\end{equation*}
The vertical line in the Figure 2 is the selected optimal separation
point for continuous data, where two curve lines are first
intercepted.
\begin{figure}[!ht]
    \caption{Determine cut-off point r.}
    \includegraphics[scale=0.5]{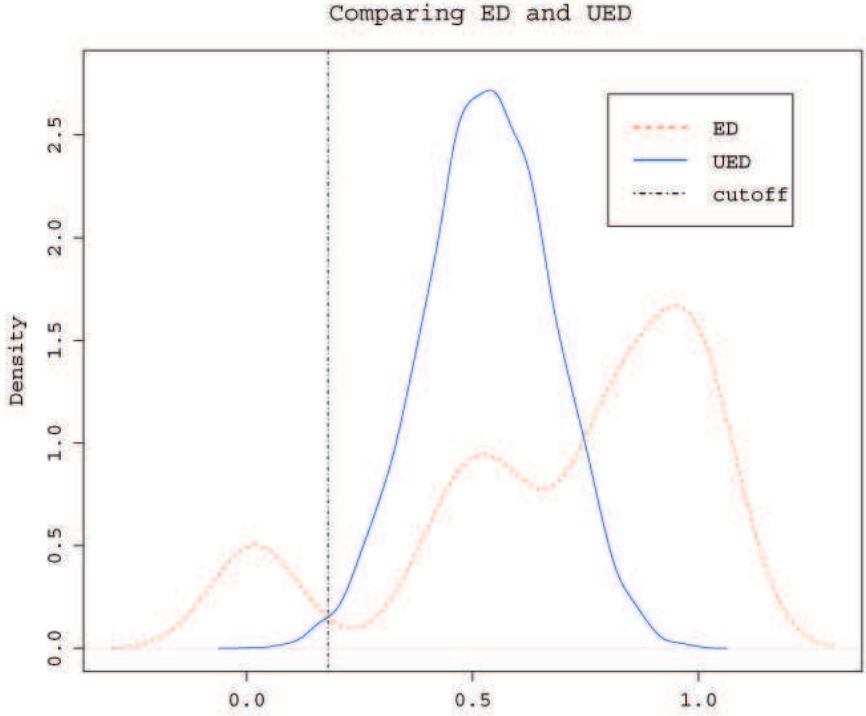}
\end{figure}

\section{Algorithm}

There are two key steps for the method. Firstly, we detect whether
there exists any statistically significant clustering pattern. We propose to use  weighted local
Chi-square test  to
determine if the observed distance vectors  differ significantly from the uniform distance vectors associated with
no cluster pattern.
Secondly, if the patterns are significant, we then extract the
clusters based on the optimal separation strategies described in the previous section.

We consider the null hypothesis $H_{0}$: There is no clustering
pattern in data set. The weighted local Chi-squared $\chi^{2*}_{w}$
is defined as:
\begin{equation*}
    \chi^{2}_{w}(r^*; (\textbf{S},\textbf{T}))= \frac{\chi^{2}_{c}(r_{c}^*;(\textbf{S},\textbf{T}))}{p}
    +\frac{\chi^{2}_{d}(r_{d}^*;(\textbf{S},\textbf{T}))}{q}, \;\;\;(\textbf{S},\textbf{T}) \in
    \Omega_{p \otimes q},
\end{equation*}
where the categorical part $\chi^{2}_{c}(r_{c}^*;\textbf{S})$ takes
form as:
\begin{equation}
  \chi^{2}_{c}(r^*;\textbf{S})  = \sum_{j=0}^{r^*_{c}} \frac{ (U_{j}(\textbf{S}) - \varepsilon_{j} )^2}{\varepsilon_{j} }
        + \frac{(\sum_{j=0}^{r^*_{c}} U_{j}(\textbf{S}) -\sum_{j=0}^{r^*_{c}} \varepsilon_{j}
        )^2}{\sum_{j=r^*_{c}+1}^{p}\varepsilon_{j}},
\end{equation}
and the continuous part $\chi^{2}_{d}(r_{d}^*;\textbf{T})$ takes the
form:
\begin{equation*}
   \chi^{2}_{d}(r^*;\textbf{T})  = \sum_{j=1}^{r^*_{d}} \frac{ (V_{j}(\textbf{T}) - \nu_{j} )^2}{\nu_{j} }
         + \frac{(\sum_{j=1}^{r^*_{d}} V_{j}(\textbf{T}) -\sum_{j=1}^{r^*_{d}} \nu_{j}
         )^2}{\sum_{j=r^*_{d}+1}^{q}\nu_{j}},
\end{equation*}
where $p$ and $p$ are number of attributes from categorical and
continuous data respectively;

After applying the statistical test with significant result, we proceed to extract clusters by determining
cluster centers \emph{\textbf{C}} and estimate cluster radius
\emph{\textbf{R}} for mixed data. Therefore, a cluster center
$\textbf{C}$ is chosen where the $\chi^{2}_{w}$ has the maximum
value:
\begin{equation*}
    \textbf{C} = \operatorname*{arg\,max}_{(\textbf{S},\textbf{T})}
    \chi_{w}^{2}.
\end{equation*}

How to determine the cluster size is the next key step to complete
cluster extraction process. Radius is the term we define to determine the
size of a cluster.  Zhang {\it et.al}(2005) gave the definition of
radius which is the maximum distance of the data points in this
cluster to its center.  Radius is the distance at which the HD
vector has its very first local minimum.  Therefore, he defined
categorical radius $R_{c}(C)$ as:
\begin{equation*}
    R_{c}(C)= \operatorname*{mim}_ {0<j<p_{c}} \{j| U_{j}(\textbf{C}) < min (U_{j-1}(\textbf{C}), U_{j+1}(\textbf{C})) \}
    -1.
\end{equation*}
For continuous part of the data, only those values before cut-off
point are sensitive for selecting radius. Figure 3 gives empirical
CDF plot of ED values where ED values jump at certain point. The
first jump point is used as the value of continuous radius. More
specifically, during each extraction iteration, we remove those
extracted data points from the rest of clustering process in order to calculate distance between
subjects to a fixed reference position.

\begin{figure}[!ht]
    \caption{Determine Radius $R_{d}$ for continuous portion of the data}
    \includegraphics[scale=0.5]{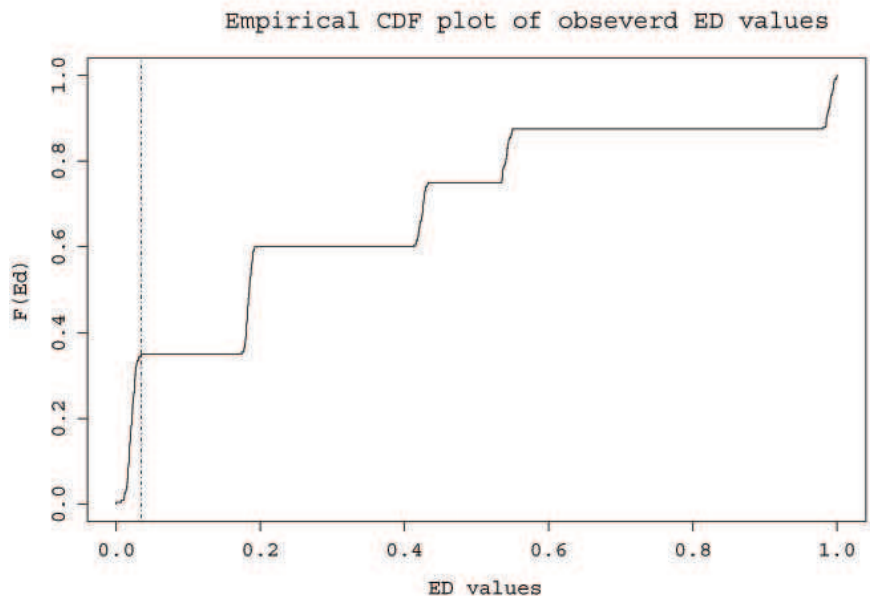}
\end{figure}

The detailed procedures for our method are described as the following:

\begin{description}
  \item[Step 1.] For each position \textit{\textbf{S}}, we calculate Hamming
  distance (HD) in the Categorical sample space $H_{c}$ and Euclidean distance (ED)
  in Continuous sample space $H_{d}$;

  \item[Step 2.] Based on HD and ED,
  calculate and compare with UHD Vector and UED Vector;

  \item[Step 3] Determine cut-off $r^{*}_{c}(S)$ and $r^{*}_{d}(S)$ for categorical and continuous data respectively; and further calculate the
  corresponding modified Chi-squared statistic $\chi^{2*}_{c}(S)$ and
  $\chi^{2*}_{d}(S)$ and obtain test statistic, weighted local
  chi-square $\chi^{2*}_{w}(S)$;

  \item[Step 4] Corresponding to the weighted local Chi-squared test, select the largest test statistic
  $\chi^{2*}_{w}(\textbf{T}^*)$; compare it with critical value $\chi^{2*}_{(0.05)}$.  If the
  max($\chi^{2*}_{w}(\textbf{T}^*)$) is smaller than $\chi^{2*}_{(0.05)}$, stop
  the algorithm; otherwise, continue to step 5;

  \item[Step 5] Assign the position who has either the largest test
  statistic $\chi^{2*}_{w}(\textbf{T}^*)$;

  \item[Step 6] Calculate categorical radius $R_{c}(\textbf{C})$ and continuous radius
  $R_{d}(\textbf{C})$;  label all data points within radius in the cluster; remove them from the current data
  set;

  \item[Step 7] Repeat Step 1 to 7 until no more significant clusters
  are detected.

\end{description}

\section{Numerical Results}
We carry out simulation studies  to examine the performances of our
proposed method. Classification rates and information gains are
calculated to compare the performance from our proposed method and
AutoClass algorithm. For simplicity, we assume all attributes are independent
in the mixed data.  The simulation setting is as the following:

\begin{enumerate}

    \item Set the number of categorical attributes $p = 10$ and each attribute takes $m_{j}$ levels which is randomly selected from the set
    $\{4,5,6\}$; Set the number of continuous attributes $q = 10$.
    \item Set the number of clusters $K_{c} = K_{d} = 3 \; or \; 5$ with various sizes.  The number of replications is $500$. Continuous case 1, 2, 3 are generated from $10$ independent normal distributions with same mean but
 variance are ranging from $0.25, 0.5$ and $1$ respectively.
    \item For categorical data, in the $k^{th}$ cluster with center
        $\textbf{C}_{k}$, generate $n_{k}$ 10-attributes vectors independently.
        More specifically,  generate for each attribute from a
        multinomial distribution with center probability $0.7$ and the
        rest probability are identically equal to $0.3/(m_{j} -1)$; For
        continuous data, $n_{k}$ 10-attributes vectors are 10 independent
        normal random variables with $\boldsymbol\mu = \textbf{C}_{k}$
        and $\sigma^2$ ranging from $0.25, 0.5$ and $1$, respectively.

\end{enumerate}
Table 1 to Table 4 provide results from the simulation experiments
with $500$ replications. Averages of classification rates (CR) and
information gains (IG) with their corresponding standard deviations
are used to evaluate two methods' performance. Table 1 to Table 4
show the results from simulated data with various settings of sample
size, number of clusters and cluster sizes.

Table 1 is obtained from
analyzing data withs with a  sample size is 200 with 3 clusters
of the sizes of 100, 75 and 25, respectively. Table 2 is obtained
from simulated data with sample size 200, cluster numbers 3 and each
cluster size 130, 45 and 25, respectively. Simulated data for Table
3 and 4 have sample size 100 and number of clusters is 5, but each
cluster size is 40, 25, 15, 10 and 10 for Table 3 and 35, 25, 20, 10
and 10 for Table 4.

It can be seen from  Table 1 that our proposed
algorithm has relatively higher classification rate and information
gain rate with lower standard deviations correspondingly, especially
in continuous portion of data.  Table 2 shows, compared to
AutoClass, our method generates significantly  higher CR with lower standard
deviations for both categorical and continuous data, for instance,
for Categorical1, AutoClass gives $75\%$ CR with $14\%$ standard
deviation, and ours gives $96\%$ CR with $5\%$ standard deviation.
Table 3 and Table 4 show the same  patterns.  In summary,
speaking, it is shown by Table 1 to 4, our proposed algorithm
consistently has higher classification rate and information gain
rate with lower standard deviation correspondingly.

\section{Discussion}
We have proposed a non-parametric clustering method based on
weighted modified Chi-sqaure test.
  Numerical results show that the proposed method
outperforms the AutoClass algorithm based on classification rate and
entropy measure for various simulation settings. The proposed  method
is most useful when neither a distance function nor a parametric model can
be assumed.
We will extend this proposed method to cluster spatial and temporal data.

\newpage
% table 1 N=200, k=3, replicate=500
\begin{table}[!ht]
 \begin{center}
 \caption{\small Average Classification Rates (CR) and Information Gains (IG) with corresponding stand deviation for each
 method. The sample size is \emph{200} with \emph{3} clusters. Each cluster has size \emph{100, 75} and \emph{25}, respectively. Replication time
 is 500.}
 \bigskip \small
\begin{tabular}{l c c |c c |c c}

  \hline\hline
  % after \\: \hline or \cline{col1-col2} \cline{col3-col4} ...
  &                           \textbf{Categ1}   & \textbf{Cont1 } & \textbf{Categ2} & \textbf{Cont2}& \textbf{Categ3 }& \textbf{Cont3}    \\
  &                           &        \footnotesize{(Var=0.25)}   &  & \footnotesize(Var=0.5)   &  & \footnotesize(Var=1)   \\
  \hline\hline

  \multicolumn{6}{l}{\textbf{Autoclass}} \\
  {\hspace{0.5cm}}$CR Mean$   & 0.9694 & 0.8278 & 0.9705 & 0.8283 & 0.9737 & 0.8421   \\
  {\hspace{0.5cm}}$CR Std$    & 0.0478 & 0.0542 & 0.0483 & 0.0545 & 0.0360 & 0.0611   \\
  {\hspace{0.5cm}}$IG Mean$   & 0.5508 & 1.0000 & 0.5787 & 1.0000 & 0.5582 & 1.0000   \\
  {\hspace{0.5cm}}$IG Std$    & 0.2201 & 0.0000 & 0.2170 & 0.0000 & 0.2142 & 0.0000   \\
  \hline

 \multicolumn{6}{l} {\textbf{Weighted local Chi-squared test}} \\
  {\hspace{0.5cm}}$CR Mean$   & 0.9728 & 0.9872 & 0.9718 & 0.9849 & 0.9743 & 0.9870   \\
  {\hspace{0.5cm}}$CR Std$    & 0.0380 & 0.0379 & 0.0458 & 0.0471 & 0.0418 & 0.0434   \\
  {\hspace{0.5cm}}$IG Mean$   & 0.8956 & 0.9683 & 0.8959 & 0.9646 & 0.9027 & 0.9706   \\
  {\hspace{0.5cm}}$IG Std$    & 0.1128 & 0.0942 & 0.1174 & 0.1044 & 0.1099 & 0.0955   \\
  \hline \\
\end{tabular}
\end{center}
\end{table}

\newpage
% table 2 N=200, k=3, replicate=500
\begin{table}[!ht]
 \begin{center}

 \caption{\small Average CR and IG with corresponding stand deviation for each
 method. The sample size is \emph{200} with \emph{3} clusters Each cluster has size \emph{130, 45} and \emph{25}, respectively. Replication is 500 times.}
 \bigskip \small
\begin{tabular}{l c c |c c |c c}

  \hline\hline
  % after \\: \hline or \cline{col1-col2} \cline{col3-col4} ...
  &                           \textbf{Categ1}   & \textbf{Cont1 } & \textbf{Categ2} & \textbf{Cont2}& \textbf{Categ3 }& \textbf{Cont3}    \\
  &                           &        \footnotesize{(Var=0.25)}   &  & \footnotesize(Var=0.5)   &  & \footnotesize(Var=1)   \\
  \hline\hline

 \multicolumn{6}{l} {\textbf{Autoclass}} \\
  {\hspace{0.5cm}}$CR Mean$   & 0.7497 & 0.6919 & 0.7530 & 0.6852 & 0.7618 & 0.6924   \\
  {\hspace{0.5cm}}$CR Std$    & 0.1375 & 0.0742 & 0.1407 & 0.0599 & 0.1414 & 0.0664   \\
  {\hspace{0.5cm}}$IG Mean$   & 0.6200 & 1.0000 & 0.6361 & 1.0000 & 0.6246 & 1.0000   \\
  {\hspace{0.5cm}}$IG Std$    & 0.2398 & 0.0000 & 0.2248 & 0.0000 & 0.2408 & 0.0000   \\
  \hline

 \multicolumn{6}{l} {\textbf{Weighted local Chi-squared test}} \\
  {\hspace{0.5cm}}$CR Mean$   & 0.9633 & 0.9742 & 0.9680 & 0.9795 & 0.9648 & 0.9762   \\
  {\hspace{0.5cm}}$CR Std$    & 0.0549 & 0.0588 & 0.0493 & 0.0531 & 0.0556 & 0.0590   \\
  {\hspace{0.5cm}}$IG Mean$   & 0.8633 & 0.9383 & 0.8764 & 0.9511 & 0.8694 & 0.9426   \\
  {\hspace{0.5cm}}$IG Std$    & 0.1585 & 0.1432 & 0.1421 & 0.1287 & 0.1584 & 0.1475   \\
  \hline \hline\\

\end{tabular}
\end{center}
\end{table}

\newpage

% table 3 N=100, k=5, replicate=500
\begin{table}[ht]
 \begin{center}
  \small
 \caption{\small Average CR and IG with corresponding standard deviations for each
 method. The sample size is \emph{100} with \emph{5} clusters. Each cluster has size \emph{40, 25, 15,10} and \emph{10},
 respectively. Replication time is 500.}
 \bigskip \small
\begin{tabular}{l c c |c c| c c}

  \hline\hline
  % after \\: \hline or \cline{col1-col2} \cline{col3-col4} ...
  &                           \textbf{Categ1}   & \textbf{Cont1 } & \textbf{Categ2} & \textbf{Cont2}& \textbf{Categ3 }& \textbf{Cont3}    \\
  &                           &        \footnotesize{(Var=0.25)}   &  & \footnotesize(Var=0.5)   &  & \footnotesize(Var=1)   \\
  \hline\hline

 \multicolumn{6}{l} {\textbf{Autoclass}} \\
  {\hspace{0.5cm}}$CR Mean$   & 0.7667 & 0.7592 & 0.7686 & 0.7268 & 0.7619 & 0.7035   \\
  {\hspace{0.5cm}}$CR Std$    & 0.0440 & 0.0403 & 0.0483 & 0.0425 & 0.0497 & 0.0371   \\
  {\hspace{0.5cm}}$IG Mean$   & 0.6423 & 0.8699 & 0.6422 & 0.8699 & 0.6390 & 0.8699   \\
  {\hspace{0.5cm}}$IG Std$    & 0.1958 & 0.0000 & 0.1938 & 0.0000 & 0.2014 & 0.0000   \\
  \hline

 \multicolumn{6}{l} {\textbf{Weighted local Chi-squared test}} \\
  {\hspace{0.5cm}}$CR Mean$   & 0.8503 & 0.8745 & 0.8507 & 0.8747 & 0.8503 & 0.8741   \\
  {\hspace{0.5cm}}$CR Std$    & 0.1002 & 0.1045 & 0.1011 & 0.1047 & 0.0998 & 0.1035   \\
  {\hspace{0.5cm}}$IG Mean$   & 0.7172 & 0.8432 & 0.7174 & 0.8460 & 0.7171 & 0.8439   \\
  {\hspace{0.5cm}}$IG Std$    & 0.1666 & 0.1474 & 0.1695 & 0.1434 & 0.1645 & 0.1438   \\
  \hline \\
\end{tabular}
\end{center}
\end{table}

\newpage
% table 4 N=100, k=5, replicate=500
\begin{table}[ht]
 \begin{center}
  \small
 \caption{\small CR and IG with corresponding stand deviations for each
 method. The sample size is \emph{100} with \emph{5} clusters.  Each cluster has size 35, 25, 20, 10 and 10, respectively.
 Replication time is 500.}
 \bigskip \small
\begin{tabular}{l c c |c c |c c}

  \hline\hline
  % after \\: \hline or \cline{col1-col2} \cline{col3-col4} ...
  &                           \textbf{Categ1}   & \textbf{Cont1 } & \textbf{Categ2} & \textbf{Cont2}& \textbf{Categ3 }& \textbf{Cont3}    \\
  &                           &        \footnotesize{(Var=0.25)}   &  & \footnotesize(Var=0.5)   &  & \footnotesize(Var=1)   \\
  \hline\hline

  \multicolumn{6}{l} {\textbf{Autoclass}} \\
  {\hspace{0.5cm}}$CR Mean$   & 0.7901 & 0.6751 & 0.7887 & 0.6532 & 0.7910 & 0.6321   \\
  {\hspace{0.5cm}}$CR Std$    & 0.0515 & 0.0338 & 0.0505 & 0.0370 & 0.0510 & 0.0346   \\
  {\hspace{0.5cm}}$IG Mean$   & 0.6728 & 0.5419 & 0.6695 & 0.5519 & 0.6813 & 0.5570   \\
  {\hspace{0.5cm}}$IG Std$    & 0.1958 & 0.0236 & 0.1923 & 0.0310 & 0.1846 & 0.0312   \\
  \hline

 \multicolumn{6}{l} {\textbf{Weighted local Chi-squared test}} \\
  {\hspace{0.5cm}}$CR Mean$   & 0.8634 & 0.8879 & 0.8581 & 0.8843 & 0.8699 & 0.8959   \\
  {\hspace{0.5cm}}$CR Std$    & 0.1009 & 0.1019 & 0.1015 & 0.1013 & 0.0999 & 0.0984   \\
  {\hspace{0.5cm}}$IG Mean$   & 0.7454 & 0.8586 & 0.7352 & 0.8536 & 0.7565 & 0.8693   \\
  {\hspace{0.5cm}}$IG Std$    & 0.1608 & 0.1384 & 0.1622 & 0.1373 & 0.1572 & 0.1286   \\
  \hline \hline\\

\end{tabular}
\end{center}
\end{table}


\newpage

\begin{thebibliography}{9}

\bibitem{Banfield}
    Banfield, J., and Raftery, A.(1993), " Model-Based Gaussian and
    Non-Gaussian Clustering," Biometrics, 49, 803-821

\bibitem{Bradley}
    Bradley, P., Fayyad, U., and Reina, c. (1998), " Scaling Clustering Algorithms to Large
    Databases," in Proceedings of the Fourth International
    conference on Knowledge Discovery and Data Mining, New York,
    August 1998, CA: AAAI Press, pp.9-15.

\bibitem{Cheeseman}
    Cheeseman, P., and Stutz, J. (1995), " Bayesian classification
    (AUTOCLASS): Theory and Results," in Advances in Knowledge
    Discovery and Data Mining, eds.  U. Fayyad, G. Piatesky-Shapiro,
    P. Smyth and R. Uthurusamy, Menlo Park, CA: AAAI Press, pp.
    153-180.

\bibitem{Fraley}
    Fraley, C., and Raftery, A. (1998), " How Many Clusters? Which Clustering Method? Answer via Model-Based Cluster
    Analysis," The Computer Journal, 41, 578-587.



\bibitem{Gasieniec}
    Gasieniec, L., jansson, J., and Lingas, A.(2004), " Approximation
    Algorithms for Hamming Clustering Problems,: Journal of Discrete
    Algorithms, 2, 289-301.

\bibitem{Gordon}
    Gordon, A. d. (1999), Classification, London: Chapman $\&$ Hall.

\bibitem{Huang}
    Huang, Z. X. (1997), " Extensions to the K-Means Algorithm for
    Clustering large Data Sets With Categorical Values," Data Mining
    and Knowledge discovery, 2, 283-304.

\bibitem{Kaufman}
    Kaufman, L., and Rousseeuw, P. J. (2005). Finding Groups in
    Data: An Introduction to Cluster Analysis, New York: Wiley.

\bibitem{Labooulais}
    Laboulais, C., Ouali, M., Le Bret, M., and Gabarro-Arpa,
    J.(2002).  Hamming Distance Geometry of a Protein Conformational
    Space. Application to the Clustering of   Molecular Dynamics Trajectory of th
    HIV-1 Intergrase Catalytic Core," Proteins: Structure, Function
    and Genetics, 47, 169-179

\bibitem{MacQueen}
    MacQueen, J. B. (1976) " Some Methods for Classification and
    Analysis of Multivariate Observations," in Proceedings of the
    5th Symposium at mathematical Statistics and Probability,
    Berkeley, CA: University of California Press, pp. 281-297.


\bibitem{Roman}
Roman, Steven. (1992). Coding and Information Theory.
Springer-Verlag, New York.


\bibitem{Zhang}
    Zhang, P., Wang, X. and Song, P.X. (2006) Clustering Categorical
    Data Based on Distance Vectors.  The Journal of the American
    Statistical Association. Vol. 101. No. 473, 355-367.

\end{thebibliography}
\end{document}